\documentclass[useAMS,usenatbib,letterpaper]{mn2e}   

\usepackage{graphicx}
\usepackage{natbib}
\usepackage{aas_macros}

\title[AA~Dor]{The mass ratio and the orbital parameters of the
sdOB binary AA~Doradus}

\author[Slavek M. Rucinski]
{Slavek M. Rucinski\thanks{E-mail: rucinski@astro.utoronto.ca}\\
Department of Astronomy, University of Toronto\\
50 St.~George St., Toronto, Ontario, Canada M5S~3H4}

\date{Accepted --.
      Received -- ;
      in original form --}

\pubyear{2009}

\begin{document}

\maketitle

\label{firstpage}

\begin{abstract}
The time sequence of 105 spectra covering one full
orbital period of AA~Dor has been analyzed.
Direct determination of $V \sin i$ for the sdOB
component from 97 spectra outside of the eclipse
for the lines Mg~II 4481 \AA\ and Si~IV 4089 \AA\
clearly indicated a substantially smaller value
than estimated before. Detailed modelling of line
profile variations for 8 spectra during the
eclipse for the Mg~II 4481 \AA\ line, combined
with the out-of-eclipse fits, gave
$V \sin i = 31.8 \pm 1.8$ km~s$^{-1}$.
The previous determinations of $V \sin i$,
based on the He~II 4686 \AA\
line, appear to be invalid because of the large natural
broadening of the line.
With the assumption of the solid-body, synchronous
rotation of the sdOB primary, the measured values of
the semi-amplitude $K_1$ and $V \sin i$
lead to the mass ratio $q = 0.213 \pm 0.013$
which in turn gives  $K_2$ and thus
the masses and radii of both components. The sdOB
component appears to be less massive
than assumed before, $M_1 = 0.25 \pm 0.05\,M_\odot$, but
the secondary has its mass--radius parameters close to
theoretically predicted for a brown dwarf,
$M_2 = 0.054 \pm 0.010\, M_\odot$ and
$R_2 = 0.089 \pm 0.005\,R_\odot$. Our results do not
agree with the recent determination of \citet{Vuc08}
based on a $K_2$ estimate from line-profile asymmetries.
\end{abstract}


\section{Introduction}
\label{intro}

AA~Doradus (LB~3459, HD~269696, $V \approx 11$)
is a well known, close, short period (0.261 d), totally eclipsing
binary consisting of a hot ($T_{eff} \approx 42,000\,K$)
sdOB, compact star and an invisible object.
A recent review of \citet{Heber2008} gives a broad
description of the sdB and sdO sub-dwarfs and of
the intense research in this field.

The eclipses of the sdOB primary component of AA~Dor
are total and deep ($\approx 0.35$ mag).
The illumination (reflection) effect from the invisible body
is very strong and highly wavelength dependent: It is
apparently mostly the heating effect as it is the strongest in
IR and almost absent in UV; the secondary eclipses are
entirely due to the eclipses of this additional, ``reflected''
light.

Previous investigations of AA~Dor
\citep{Kilk78,Kilk79,Hill96,Hill03,RW03} have
resulted in an excellent and detailed description of its
geometry in {\it relative\/} units:
The relative radii ($r_1=0.1419$, $r_2=0.0764$)
and the orbital inclination ($i=89.2\deg$) are known to a very
high accuracy. However, the {\it absolute\/} parameters of AA~Dor
have not been known until the very recent paper of \citet{Vuc08}.
This is related to the single-lined (SB1)
character of the spectrum which can provide only the semi-amplitude
$K_1$ \citep{Hill03,RW03}. Thus, the absolute sizes and
masses could only be guessed using various reasonable assumptions
on the mass of the visible star and/or the mass ratio. However,
the sdOB component is not an average and typical star
so such assumptions may be risky.

\citet{Vuc08}, through ingenious data processing of the
same data as used in this paper, have been
able to detect weak asymmetries in the hydrogen line wide wings.
The asymmetric components were in emission and
moved in anti-phase to the primary component motion permitting
an estimate of $K_2$. In this paper we show that this
determination does not agree with our entirely independent analysis
based on the measured value of $V \sin i$ and the
assumptions of the solid-body, synchronous rotation for the
primary component.

\begin{figure}
\begin{center}
\includegraphics[width=70mm]{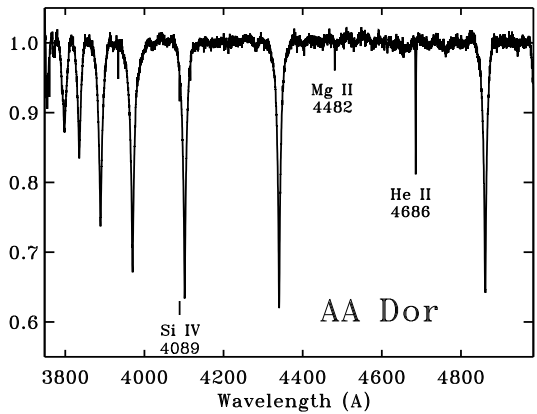}
\caption{The first of 105 spectra at the orbital
phase 0.494, used in this analysis and taken here as an example.
The three spectral lines used in this paper for
$V \sin i$ determination are marked.}
\label{fig1}
\end{center}
\end{figure}

\section{Observational data}
\label{obs}

The analysis presented here is based on the
collection of 105 spectra obtained within the program of
\citet{RW03} on January 8, 2001 using the 8m ESO telescope
and the UVES spectrograph. The observations became available
to the author through the kindness of Dr.\ T.\ Rauch;
they are  accessible through the European
Southern Observatory (ESO) archives.

The spectra were obtained in a somewhat cloudy weather
(roughly 50\% changes in the air transparency),
over slightly more than one orbital cycle of the star,
starting at orbital phases of the
secondary eclipse (the hot star in front).
The exposures were taken at equal intervals of 3.75 minutes.
The nominal resolving power was $R \approx 48,000$,
corresponding to 6.2 km~s$^{-1}$ (for
a Gaussian profile, $\sigma_{instr} \approx 3.0$ km~s$^{-1}$).
The wavelength sampling of the spectra was  0.015 \AA\,
i.e.\ about 1 km~s$^{-1}$. Although the AA~Dor spectra
have been fully reduced through the ESO pipeline and expressed
in the heliocentric radial velocity system, no standard
or calibration star spectra were provided.
The remaining details are given in \citet{RW03}.

Figure~\ref{fig1} shows the rectified spectrum AA~Dor, the
first of the 105 used; the average spectrum of
AA~Dor formed from the same observations is shown
magnified in Figure~2 of  \citet{RW03}. The latter figure
was used as a guidance during rectification of the spectra because
the raw data showed relatively large, wave-like intensity
variations of the order of 50\% along the covered
wavelength range.

Of note in Figure~\ref{fig1} are the very wide hydrogen
lines extending in radial velocities over a few thousand km/s.
The only obviously sharp and
relatively deep (about 20~\% depth) line
is the He~II 4686 \AA\ line. This line was used
by \citet{RW03} for a detailed modelling of the rotational
broadening; however the intrinsic factors of the
fine structure splitting and of a large thermal broadening
(light atoms) gave only moderate quality results for the
important quantity of the projected rotational velocity
$V \sin i$. For that reason, we used much shallower but
sharper lines of Mg~II 4482 \AA\ and Si~IV 4089 \AA\ in our
determination of $V \sin i$ (Sections~\ref{out-of-ecl} --
\ref{observed}).

The extensive analysis of the ESO/UVES spectral
series by \citet{RW03},
while consistent with that of \citet{Hill03},
has not improved our knowledge of
the system beyond providing the well defined value of
$K_1 = 39.19 \pm 0.05$ km~s$^{-1}$ and
a much less well defined
$V_{rot} = 47 \pm 5$ km~s$^{-1}$ (note that the latter
quantity has the error one hundred times
larger than the former). It is not
clear if the latter value is the aspect-corrected one;
\citet{Hill03} used $V \sin i = 43 \pm 5$ km~s$^{-1}$.
Assuming a plausible range of the masses for the primary,
$M_1 \approx 0.33 - 0.47\,M_\odot$,
and the value of $K_1$ as above, \citet{Hill03} arrived at
$M_2 \approx 0.064 - 0.082\,M_\odot$ for the invisible object.
While this was the best that could be derived, it definitely
requires confirmation. We show in this paper that a targeted
determination of $V \sin i$ may indicate a rather different
set of absolute parameters of AA~Dor.

Recently, after the analysis for this paper was concluded,
a new determination of the rotational velocity of AA~Dor
appeared \citep{Fleig08}. It is based on a detailed analysis of
far ultraviolet spectra from the FUSE satellite.
The new value is substantially smaller than used
before, $V_{rot} = 35 \pm 5$ km~s$^{-1}$, and
better agrees with the results of this paper. However, our
analysis uses a very different approach and results in a smaller
uncertainty so it is useful as an entirely independent
determination.

\section{Rotation velocity of the primary component and
the parameters of the system}
\label{theor}

The eclipsing binary system AA~Dor has an excellent light-curve
solution and the relative sizes of components are very well
known \citep{Hill03}.
Yet, the physical scale of the system, which is needed for
derivation of the masses requires not only the easily
measurable semi-amplitude $K_1$ but also information
on the motion of the secondary component. Although an
estimate of $K_2$ has been recently found \citep{Vuc08},
it may be biased as it was derived from tiny spectral-line
asymmetries. Instead, we propose to utilize as the
second quantity the value of the projected equatorial
velocity of the primary star, $V \sin i$.
One can then estimate the mass ratio
of the system, $q = M_2/M_1$, and obtain a full
description of the binary system in {\it physical units\/}
by making the two assumptions about the primary component:
(1)~the solid-body rotation and
(2)~the perfect rotation--orbital motion synchronism.
The ratio of the projected equatorial velocity to the observed
orbital semi-amplitude,
\begin{equation}
V \sin i/K_1 = r_1 (1 + 1/q)
\label{eq1}
\end{equation}
relates the relative size of the visible star ($r_1$) to the
relative size of its orbit. This can be used to determine
$q$ and then $K_2 = K_1/q$.

While $K_1$ is relatively easy to find for AA~Dor,
$V \sin i$ has been poorly determined. We attempt to
determine $V \sin i$ first by direct fits to the
spectral line profiles (Section~\ref{out-of-ecl})
and then by modelling of the line profile variations
during eclipses of the primary (Section~\ref{eclipse}).

\begin{figure}
\begin{center}
\includegraphics[width=80mm]{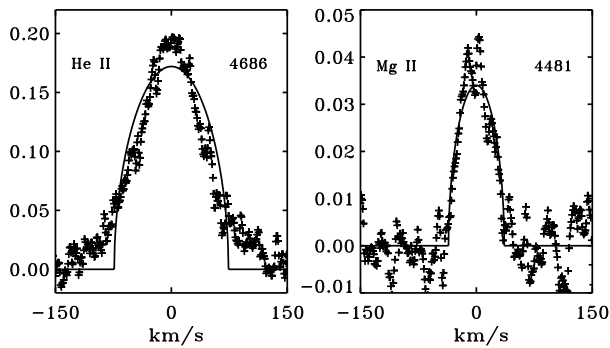}
\caption{Two spectral lines of He~II 4686 \AA\
and Mg~II 4481 \AA, in the first spectrum
of the series of 105 spectra, at the orbital
phase 0.494 (the primary star in front).
The rotational profile fits are for mean parameters
established for all 97 out-of-eclipse spectra
which were used for determinations of $K_1$
and $V \sin i$. The line profiles are shown
inverted with the vertical scale representing the
depth of the line relative the local continuum.
}
\label{fig2}
\end{center}
\end{figure}

\section{Spectral line profile rotational fits
outside of eclipses}
\label{out-of-ecl}

There are very few sharp lines in the spectrum
of AA~Dor (the full spectrum of AA~Dor is
shown in Figure~2 of \citet{RW03}).
The only visibly sharp and
deep (20\%) line is the He~II 4686 \AA\ line.
The lines of Mg~II 4481 \AA\ and Si~IV 4089 \AA\
have depths of about 4 -- 5\%. Other sharp lines are
either interstellar or appear in the UV part of
the spectrum where their groups would require
detailed modelling of the mutual overlap and blending.

The He~II 4686 \AA\ line was modelled in
detail by \citet{RW03} providing determinations
of $K_1$ and of $V \sin i$. However, the line does not look
rotationally broadened at all (Figure~\ref{fig2}) so
direct rotation profile fits are expected to give biased
results. In spite of this, all 97 out-of-eclipse
line profiles for this line were fitted by by the
standard rotation profile (assuming the limb darkening of
$u = 0.16$ which is appropriate for the high
temperature of the star) to provide
orbital velocities of the visible star and
repeated determinations of $V \sin i$.
The orbital solution for a circular orbit is:
$K_1 = 39.41 \pm 0.13$ km~s$^{-1}$,
$T_0 = 2451917.15400 \pm 0.00016$. Because there were
no standard star observations to relate to, the
centre of mass velocity was determined but
later adjusted to zero by the choice of
the reference wavelength, $V_0 = 0 \pm 0.09$ km~s$^{-1}$.
The mean standard errors have been determined
by the bootstrap sampling and are more realistic than ones
given by the formal least-squares linear estimates.
For the eclipse profile fits, we used a slightly different
$T_0 = 2451917.15389 \pm 0.00019$ which is based on the weighted
mean of the orbital solutions for the He~II line
and the Mg~II line (see below).
Relative to the values given by \citet{RW03},
the adopted $K_1$ is slightly larger by 0.2 km~s$^{-1}$
but has a four times larger (and probably
more realistic) error. The value of $T_0$
is shifted forward by 0.0012 d. The variable $T_0$
was included in the solution to account for an
(unlikely) chance of spots or other asymmetries
manifesting themselves only spectroscopically;
obviously, an addition of one additional degree of
freedom can only lower the quality of $V \sin i$
determination, but was done to stay conservative
with the determination.

The value of the formally obtained rotation profile
width of the He~II line is $V \sin i = 74.9$ km~s$^{-1}$.
The error of the single determination is 1.5 km~s$^{-1}$
and error of the mean of 97 determinations is some
ten times smaller; however, such an error of the mean estimate
would not make allowance for the systematic departures
of the fit and would reflect only the consistently same shape of the
line profile at all phases. Thus, we do not attach
too much significance to the formal errors and note
that the determination of $V \sin i$ is large; it
obviously combines
genuine rotation with other sources of broadening.
The spectrograph resolution is characterized by
$\sigma \approx 3$ km~s$^{-1}$ so that the dominant effect
must come from the natural broadening of the He~II line.
The He ions are light so that for the
temperature of the AA~Dor primary, the Doppler
thermal widths is expected to be
relatively large, $\xi_0 \approx 16$ km~s$^{-1}$.
The line additionally shows multiple components of
the fine structure splitting, as discussed in \citet{RW03},
but even these two effects combined are not large enough.
This indicates that the He~II 4686 \AA\ line, in spite of its sharp
appearance -- at least compared with the hydrogen lines --
is not sharp enough for a $V \sin i$ determination.

The Mg~II 4481 \AA\ and Si~IV 4089 \AA\ lines are
expected to be sharper than the helium line
because their thermal Doppler broadening
is $\xi_0 \approx 5.5 - 6$ km~s$^{-1}$.
When combined with the spectrograph resolution, we may expect
their natural profile to have a width of
$\approx 7 - 8$ km~s$^{-1}$. The Mg~II 4481 \AA\ line
is indeed much sharper than the He~II line
(Figure~\ref{fig2}) and its broadening does appear to be
dominated by rotation. Fits of the rotational
profile to the line profiles outside of the eclipse gave
$K_1 = 39.6 \pm 0.33$ km~s$^{-1}$ and
$V \sin i = 36.1$ km~s$^{-1}$; for the latter quantity
the formal error of a single determination
is 3.4 km~s$^{-1}$ and does not
include any systematic deviations in the $V \sin i$ fits.

The third of the considered lines,
Si~IV 4089 \AA, is located on the wide and curved wing of the
H-$\delta$ line. The raw spectra had
a sensitivity dip in this region
by some 1/3 of the highest counts with an associated
increase in the noise. The line is also slightly shallower than
the Mg~II 4481 \AA\ line.
For these reasons we would consider results for this line
as less reliable than those for the Mg~II 4481 \AA\ line.
Individual fits of the rotational profiles to all
out-of-eclipse Si~IV line profiles gave
$V \sin i = 31.9$ km~s$^{-1}$, with the
formal error of a single determination of 2.4 km~s$^{-1}$.
The small value of $V \sin i$ may be partly due to the
distortion of the baseline by the H-$\delta$ line wing,
but indicates that this quantity cannot be large.

To summarize: The value of $V \sin i$ estimated from direct
line-profile fits to the Mg~II 4481 \AA\ and Si~IV 4089 \AA\
lines is $36.1$ and $31.9$ km~s$^{-1}$, respectively, with the
latter determination much less reliable than the former.
The fits for the He~II 4686 \AA\ line gave entirely unreliable
results apparently due to strong natural broadening
of the line. In the next sections, we will address the matter
of separation of the natural broadening contribution from
the $V \sin i$ estimates through utilization of the line-profile
variations during eclipses.

\begin{figure}
\begin{center}
\includegraphics[width=80mm]{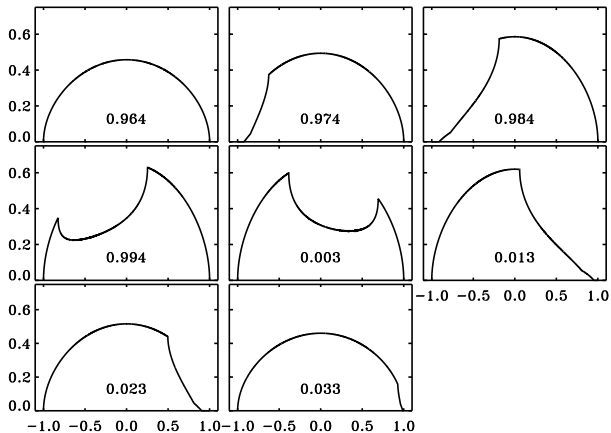}
\caption{Predicted eclipse variations for an infinitely
narrow spectral line in the spectrum of
AA~Dor during the primary eclipse.
The X-axis is expressed in units of
$V \sin i$ while the Y-axis is in arbitrary
units but with the profiles normalized to give
the same integral for all phases.
The profiles have been calculated
for eight observed phases as given by numbers
in each panel.
}
\label{fig3}
\end{center}
\end{figure}

\section{Predicted spectral line-profile variations during
eclipses}
\label{eclipse}

For determination of $V \sin i$, one can utilize
the spectral line profile variations as the
secondary component of AA~Dor
scans the velocity field of the eclipsed sdOB
component. Because the geometry of the binary is fully
defined in the relative units of the mean separation,
the line profile variations are relatively easy to predict
for the case of the eclipsed body rotating as a solid-body,
rotationally-synchronized star.

The theoretical line profile variations during the eclipse
(profiles called $R$ below) calculated by direct
integration over the uneclipsed portion of the primary star disk
are shown for 8 eclipse phases of AA~Dor in Figure~\ref{fig3}.
The X-axis in Figure~\ref{fig3}
is in units of $V \sin i$ which is the free
parameter of the problem. The profiles were computed
for the limb darkening coefficient of $u = 0.16$ and then
normalized to the total intensity of
unity. This normalization is appropriate when the continuum
for each phases is rectified and normalized to unity.
The line profile variations are predicted to be enormous,
a fact which may be used to determine
the value of $V \sin i$. Such
a determination would be independent of the out-of-eclipse
line-shape fits which by necessity combine two similarly
acting contributions of rotation and of natural line broadening.

The observed profile, $S (v)$, can be modelled by assuming
that it is a convolution of the
theoretical eclipse profile, $R (v - v_c; V \sin i)$
(such as in Figure~\ref{fig3}), but shifted by the predicted
orbital velocity of the primary mass centre $v_c$ and
stretched with the X-axis scale depending on $V \sin i$)
with a natural-broadening profile, $N (v; p)$:
\begin{equation}
S (v) = \int R(v - v_c; V \sin i) \, N(v; p) \, dv
\label{eq2}
\end{equation}
The natural broadening is here meant to signify
any contributions not related to rotation and thus
not sensitive to eclipses, such
as a combination of the Doppler thermal
broadening with the spectrograph finite resolution.
In the equation above, the integration variable is the
radial velocity, $v$, while $p$ is for any parameters used to
describe the natural line broadening. The natural
broadening can be assumed, for example, to be
the velocity Doppler broadening
$N \propto e^{-(v/\xi_0)^2}$, with the
dispersion linked to the effective temperature
$T_{eff}$ and the atomic mass of the ion $m$ through
$\xi_0 = \sqrt{2 \mathcal{R} T_{eff}/m}$.

Fits of the $S(v)$ profiles to the observed ones must be done
for all eclipse phases simultaneously, with (1)~the
same profile $N$ (given explicitly or described
by, say, a single parameter of
the Doppler width), (2)~a value of $V \sin i$ and
(3)~one scaling factor (when $R$ is normalized to
unit strength). With only three free parameters,
the problem is a relatively simple one
in terms of the parametric description, but somewhat complex
for a solution because of the non-linear
properties of the convolution operation.
We describe below results of least-squares, non-linear
fits for two spectral lines of He~II 4686 \AA\ and Mg~II 4482 \AA.
Variations of the Si~IV 4089 \AA\ line were not modelled because of
the curved baseline around this line caused by the H$\delta$ wings
and by the generally poorer definition of the spectrum in this
region.

\section{Observed eclipse variations of the
spectral line profiles}
\label{observed}

\begin{figure}
\begin{center}
\includegraphics[width=80mm]{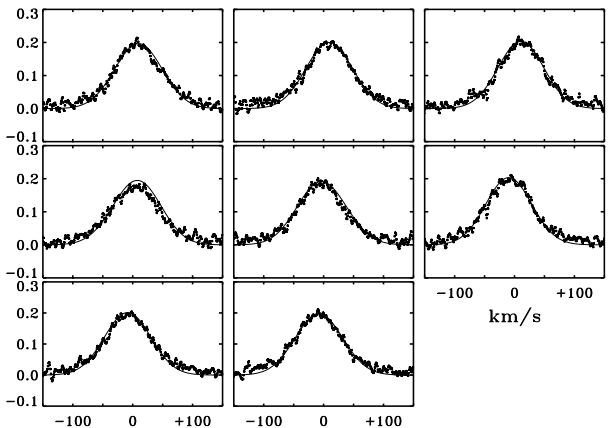}
\caption{The line profile fits of the eclipse model
to the He~II 4686 \AA\ line, for the same eight phases as
in Figure~\ref{fig3}.
The global fit utilized only three free parameters
(the best values in parentheses):
(1)~the natural, Gaussian broadening profile
($\xi_0 = 50$ km~s$^{-1}$), (2)~$V \sin i$ (35 km~s$^{-1}$)
and (3)~a common scaling factor.
The X-axis is in the observed radial velocities
shifted to an arbitrary zero point while the
Y-axis gives the inverted depth of the line.
}
\label{fig4}
\end{center}
\end{figure}

The He~II 4686 \AA\ line profile variations during
the eclipse (Figure~\ref{fig4}) are very unlike what
is expected (Figure~\ref{fig3}). This
confirms that the natural broadening
of this line totally dominates over the
rotational broadening. The eclipse variations are almost
invisible and the observed line -- except for small centroid
shifts and some changes in the overall width --
looks practically the same at all eclipse phases.
In a least-squares solution, the broadening profile had a
tendency to evolve to a large value of
$\xi_0 \approx 50$ km~s$^{-1}$, with a very large
uncertainty of $\pm 8$ km~s$^{-1}$.
For such a choice, the best
value of the rotational broadening is
$V \sin i = 35$ km~s$^{-1}$. The fits are reasonably good,
but (1)~because the eclipse effects
are basically not present, the two broadenings
simply combine (as they do outside of the eclipses);
(2)~the estimates of $\xi_0$ and $V \sin i$
are strongly anti-correlated
with each determined to at best $\pm 8$ km~s$^{-1}$.
We note that we have no explanation for the large width of
the natural profile.
This result casts a doubt on the previous determination
of $V \sin i = 43 \pm 5$ km~s$^{-1}$
utilizing the He~II 4686 \AA\ line \citep{RW03}
and points at a need to use a narrower line.
However, it agrees well with the recent determination
of \citet{Fleig08} of $35 \pm 5$ km~s$^{-1}$ from the
FUSE data.

\begin{figure}
\begin{center}
\includegraphics[width=80mm]{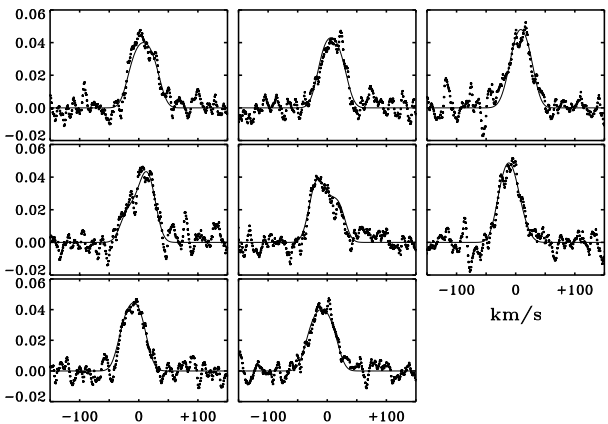}
\caption{The line profile fits of our model
to the Mg~II 4481 \AA\ line
for the eight phases during the eclipse, the same
as in Figures~\ref{fig3} and \ref{fig4}.
The derived parameters are $V \sin i = 31.8$ km~s$^{-1}$
and the natural Gaussian broadening with
$\xi_0 = 13.8$ km~s$^{-1}$.
Note that the vertical scale is much expanded relative
to Figure~\ref{fig4} so that the noise is strongly
visible, but that the horizontal scale is the same
directly showing the lesser broadening of the
Mg~II 4481 \AA\ line. Note also the detailed changes of
this line, particularly for the
central eclipse phases (panels 4 and 5).
}
\label{fig5}
\end{center}
\end{figure}

The Mg~II 4481 \AA\ line (Figure~\ref{fig5} gave the
most reliable results. The linearized least-squares
fits resulted in the Doppler width $\xi_0 = 13.8 \pm 0.4$
km~s$^{-1}$ and $V \sin i = 31.8 \pm 0.4$ km~s$^{-1}$.
The error estimates were obtained through bootstrap
experiments over the whole ensemble of several thousand of
data points; separate experiments utilizing re-sampling
at the level of whole individual
spectra suggest larger errors, $\approx 1.5$ km~s$^{-1}$.
We note that the natural width is again
substantially larger than expected.

The adopted $V \sin i$ for the visible star, determined
as a weighted mean of our determinations, is
$V \sin i = 31.8 \pm 1.8$ km~s$^{-1}$. The estimate
of the uncertainty takes into account the fact that
the out-of-eclipse $V \sin i$ determinations in
Section~\ref{out-of-ecl} are biased by the
contribution of the instrumental as well as of the
uncertain natural broadening; accordingly, they were
given a low weight. We note that the values of $\xi_0$
that we were able to estimate from the line-profile
eclipse variations turned larger than expected.

The value of $V \sin i$ is
reduced by about one quarter relative to the previous
determination of \citet{RW03} (43 km~s$^{-1}$)
and is even smaller than the more recent
determination of \citet{Fleig08} (35 km~s$^{-1}$).
This -- together with
the assumptions of the solid body, synchronous rotation --
leads to different orbital parameters of AA~Dor than
estimated before. This is discussed below.

\begin{table}
\begin{small}
\caption{Parameters of AA~Dor
\label{tab_param}}
\begin{center}
\begin{tabular}{llcc}
\hline
Parameter & Unit & Value & Std Error \\
\hline
$P$                  & day         & 0.261582   & fixed \\
$i$                  & degree      & 89.2       & fixed \\
$V \sin i$           & km~s$^{-1}$ & 31.8       & 1.8   \\
$K_1$                & km~s$^{-1}$ & 39.4       & 0.2   \\
$K_2$                & km~s$^{-1}$ & 184.7      & 12.3  \\
$q = M_2/M_1$        &             & 0.213      & 0.013 \\
$(M_1+M_2) \sin^3 i$ & $M_\odot$   & 0.305      & 0.055 \\
$M_1$                & $M_\odot$   & 0.251      & 0.046 \\
$M_2$                & $M_\odot$   & 0.054      & 0.010 \\
$a \sin i$           & $R_\odot$   & 1.158      & 0.070 \\
$R_1$                & $R_\odot$   & 0.165      & 0.009 \\
$R_2$                & $R_\odot$   & 0.089      & 0.005 \\
\hline
\end{tabular}
\end{center}
\end{small}
\end{table}

\begin{figure}
\begin{center}
\includegraphics[width=70mm]{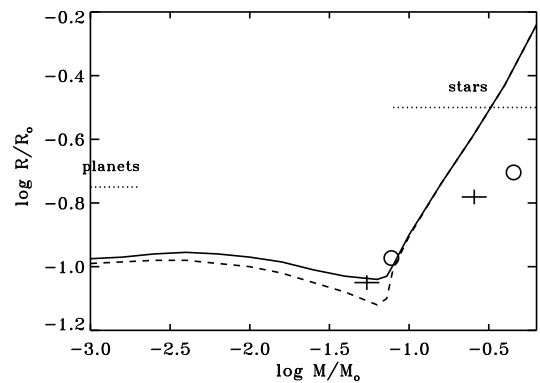}
\caption{The mass -- radius relation for the solar
composition (ages 1 Gyr and 10 Gyr) copied
from \citet{Chab08}, with the two components of
AA~Dor marked by symbols. The crosses representing
{\it formal\/} errors are for our determination
while the circles are for the determination of
\citet{Vuc08}.
The dotted lines give the customary mass ranges for low-mass
stars and massive planets.
As expected, the sdOB primary is smaller
than Main Sequence Stars, while the secondary appears to
have parameters of an object at the border
between brown dwarfs and low mass stars; the two
solutions place it at both sides of the border.
}
\label{fig6}
\end{center}
\end{figure}

\section{Parameters of AA~Dor}
\label{para}

Having the value of $V \sin i$, we can determine the mass
ratio $q$ from Eq.~\ref{eq1} and then $K_2$. With the assumed
inclination of the orbit from \citet{Hill03},
the resulting parameters of the binary system are as
listed in Table~\ref{tab_param}.
Note that in Table~\ref{tab_param}, we assumed a four times
larger $K_1$ error than in the previous solutions,
in accordance with the spread in the
solutions based on the same data, as given
in this paper and in \citet{RW03}. This way,
the disparity of errors of $K_1$ and $V \sin i$ appears
to be more acceptable than in \citet{RW03} where they differed by
a factor of one hundred times (0.05 and 5 km~s$^{-1}$,
respectively).

The solution as in Table~\ref{tab_param}
gives the mass of the primary sdOB component,
$M_1 = 0.25 \pm 0.05\,M_\odot$, well outside the previously
assumed and deemed plausible range of $0.33 - 0.47 M_\odot$
\citep{Hill03}. This is not entirely surprising because
the primary of AA~Dor is thought to be a core of
a much more massive star which underwent stripping in
a common-envelope episode.

In contrast to the primary, the parameters of
the secondary component are surprisingly
``normal'': With $M_2 = 0.054 \pm 0.010\, M_\odot$
and $R_2 = 0.089 \pm 0.005\,R_\odot$,
this object appears to fully
obey the mass -- radius relation expected for brown dwarfs.
In the recent discussion of the mass--radius relation
by \citep{Chab08} only two objects with well determined
parameters happen to be located
in the domain between the Main Sequence stars
and Jupiter-like planets: Hat--P--2b and COROT--3b.
The secondary of AA~Dor appears at the very minimum
of the radius in this relation, see Figure~\ref{fig6}.

While we are satisfied by the results, there is a problem:
They do not agree with the new determination of \citet{Vuc08}.
Their direct estimate of $K_2 = 230$ km~s$^{-1}$ is substantially
larger than our $K_2 = 185$ km~s$^{-1}$ so that both
masses are larger than ours ($M_1 = 0.45\,M_\odot$ and
$M_2 = 0.076\,M_\odot$). It is hard to imagine that
the value derived by \citet{Vuc08} is in some way biased by
the illumination effect as then one would actually expect
a {\it reduction\/} in $K_2$; however, the way how $K_2$ was
estimated -- through small line-profile
asymmetries -- does leave a wide margin of systematic
uncertainty.

Our determination of the absolute parameters of
AA~Dor totally depends on the strong assumptions of
the solid-body, synchronous
rotation of the primary component. If any of these
are not fulfilled, then our analysis is entirely invalid.
The solid-body rotation assumption is --
in principle -- verifiable
through analysis of the line-profile variations, but the
current data are not accurate enough. We note that
with the commonly agreed $K_1$ and with the new
$K_2$ by \citet{Vuc08}, the expected
$V \sin i = 38$ km~s$^{-1}$, and this is entirely ruled out
by our analysis. So, either our assumptions on the
primary component rotation are invalid or the determination
of \citet{Vuc08} is biased. The case of AA~Dor is still not
closed.

\medskip

The author would like to thank Dr.\ Thomas Rauch for
providing the data and for important advice on the
recent literature, Dr.\ Theo Pribulla for illuminating
discussions, Dr. Ron Hilditch
for reading of the draft of the paper and his
excellent comments and the reviewer
of the first version of the paper for detailed
suggestions.

Research support from the Natural Sciences and Engineering
Council of Canada is acknowledged with gratitude.

\end{document}